\documentclass[12pt]{iopart}

\usepackage{graphicx}
\usepackage{epsfig}
\usepackage{citesort}
\usepackage{xspace}
\usepackage{color}
\usepackage{verbatim}

\newcommand{\text}{\rm }

\newcommand{\grtsim}{\,\rlap{\lower3.7pt\hbox{$\mathchar\sim$}}
\raise1pt\hbox{$>$}\,}
\newcommand{\lesssim}{\,\rlap{\lower3.7pt\hbox{$\mathchar\sim$}}
\raise1pt\hbox{$<$}\,}

\begin{document}

\title[Search for 14.4 keV solar axions with CAST]
      {Search for 14.4 keV solar axions emitted in the M1-transition of $^{57}$Fe nuclei with CAST}

\newcommand{\CERN}{European Organization for Nuclear Research (CERN),
CH-1211 Gen\`eve 23, Switzerland}
\newcommand{\Saclay}{IRFU, Centre d'\'Etudes Nucl\'eaires de Saclay,
Gif-sur-Yvette, France}

\newcommand{\Darmstadt}{Technische Universit\"at Darmstadt,
    Institut f\"{u}r Kernphysik, Schlossgartenstrasse 9, 64289 Darmstadt,
  Germany}

\newcommand{\MPE}{Max-Planck-Institut f\"{u}r extraterrestrische
  Physik, Giessenbachstrasse, 85748 Garching, Germany}

\newcommand{\Zaragoza}{Instituto de F\'{\i}sica Nuclear y Altas Energ\'{\i}as,
Universidad de Zaragoza, Zaragoza, Spain }
\newcommand{\Chicago}{Enrico Fermi Institute and KICP, University of Chicago,
Chicago, IL, USA}
\newcommand{\Thessaloniki}{Aristotle University of Thessaloniki, Thessaloniki,
Greece}
\newcommand{\Athens}{National Center for Scientific Research ``Demokritos'',
Athens, Greece}
\newcommand{\Freiburg}{Albert-Ludwigs-Universit\"{a}t Freiburg, Freiburg,
Germany}
\newcommand{\INR}{Institute for Nuclear Research, Russian Academy of
Sciences, Moscow, Russia}
\newcommand{\Vancouver}{Department of Physics and Astronomy, University of
British Columbia, Vancouver, Canada }
\newcommand{\Frankfurt}{Johann Wolfgang Goethe-Universit\"at, Institut f\"ur
Angewandte Physik, Frankfurt am Main, Germany}
\newcommand{\MPI}{Max-Planck-Institut f\"{u}r Physik
(Werner-Heisenberg-Institut), F\"ohringer Ring 6, 80805 M\"unchen, Germany}
\newcommand{\Zagreb}{Rudjer Bo\v{s}kovi\'{c} Institute,
Bijeni\v{c}ka cesta 54, P.O.Box 180, HR-10002 Zagreb, Croatia}
\newcommand{\Pisa}{Scuola Normale Superiore, Pisa, Italy}
\newcommand{\Edmonton}{Department of Physics, University of Alberta, Edmonton,
             T6G2G7, Canada}
\newcommand{\BNL}{Brookhaven National Laboratory, NY, USA}
\newcommand{\Sac}{DAPNIA, CEA-Saclay, Gif-sur-Yvette, France}
\newcommand{\Patras}{University of Patras, Patras, Greece}
\newcommand{\Lyon}{Inst. de Physique Nucl\'eaire, Lyon, France}
\newcommand{\PAC}{Particle Astrophysics Center - Fermi National Accelerator
                  Laboratory, Batavia, IL~60510, USA}
\newcommand{\Bochum}{Ruhr-Universit\"{a}t Bochum, Bochum, Germany}
\newcommand{\Karlsruhe}{Institut f\"{u}r Experimentelle Kernphysik, Universit\"{a}t Karlsruhe, Karlsruhe, Germany}
\newcommand{\Fermilab}{Fermi National Accelerator Laboratory, Batavia, IL, USA}
\newcommand{\StanfordSLAC}{Stanford University and SLAC National Accelerator Laboratory, Stanford, CA, USA}

\author{S~Andriamonje$^{2}$, S~Aune$^{2}$,
D~Autiero$^{1,16}$,
K~Barth$^{1}$, A~Belov$^{11}$, B~Beltr\'an$^{6,17}$,
H~Br\"auninger$^{5}$, J~M~Carmona$^{6}$, S~Cebri\'an$^{6}$,
J~I~Collar$^{7}$, T~Dafni$^{2,4,18}$,
M~Davenport$^{1}$, L~Di~Lella$^{1,19}$,
C~Eleftheriadis$^{8}$, J~Englhauser$^{5}$,
G~Fanourakis$^{9}$, E~Ferrer-Ribas$^{2}$,
H~Fischer$^{10}$, J~Franz$^{10}$, P~Friedrich$^{5}$, T~Geralis$^{9}$,
I~Giomataris$^{2}$, S~Gninenko$^{11}$, H~G\'omez$^{6}$,
M~Hasinoff$^{12}$, F~H~Heinsius$^{10}$,
D~H~H~Hoffmann$^{4}$, I~G~Irastorza$^{2,6}$, J~Jacoby$^{13}$,
K~Jakov\v{c}i\'{c}$^{15}$, D~Kang$^{10}$,
K~K\"onigsmann$^{10}$,
R~Kotthaus$^{14}$, M~Kr\v{c}mar$^{15}$,
K~Kousouris$^{9}$,
M~Kuster$^{4,5}$, B~Laki\'{c}$^{15}$, C~Lasseur$^{1}$,
A~Liolios$^{8}$, A~Ljubi\v{c}i\'{c}$^{15}$, G~Lutz$^{14}$, G~Luz\'on$^{6}$,
D~Miller$^{7,20}$,
 J~Morales$^{6,21}$, A~Ortiz$^{6}$,
T~Papaevangelou$^{1,2}$,
A~Placci$^{1}$, G~Raffelt$^{14}$,
H~Riege$^{4}$, A~Rodr\'iguez$^{6}$, J~Ruz$^{6}$, I~Savvidis$^{8}$,
Y~Semertzidis$^{3,22}$,
P~Serpico$^{1,14}$,
L~Stewart$^{1}$, J~Vieira$^{7}$, J~Villar$^{6}$, J~Vogel$^{10}$,
L~Walckiers$^{1}$ and K~Zioutas$^{1,3}$\\
(CAST Collaboration)}

\address{$^{1}$\CERN}
\address{$^{2}$\Saclay}
\address{$^{3}$\Patras}
\address{$^{4}$\Darmstadt}
\address{$^{5}$\MPE}
\address{$^{6}$\Zaragoza}
\address{$^{7}$\Chicago}
\address{$^{8}$\Thessaloniki}
\address{$^{9}$\Athens}
\address{$^{10}$\Freiburg}
\address{$^{11}$\INR}
\address{$^{12}$\Vancouver}
\address{$^{13}$\Frankfurt}
\address{$^{14}$\MPI}
\address{$^{15}$\Zagreb}
\address{$^{16}$Present address: \Lyon}
\address{$^{17}$Present address: \Edmonton}
\address{$^{18}$Present address: \Zaragoza}
\address{$^{19}$Present address: \Pisa}
\address{$^{20}$Present address: \StanfordSLAC}
\address{$^{21}$Deceased}
\address{$^{22}$Permanent address: \BNL}

\ead{Kresimir.Jakovcic@irb.hr}


\begin{abstract}
  We have searched for 14.4~keV solar axions or more general axion-like particles (ALPs), that may be emitted in the M1 nuclear transition of $^{57}$Fe, by using the axion-to-photon conversion in the CERN Axion Solar Telescope (CAST) with evacuated magnet bores (Phase I). From the absence of excess of the monoenergetic X-rays when the magnet was pointing to the Sun, we set model-independent constraints on the coupling constants of
  pseudoscalar particles that couple to two photons and to a nucleon $g_{\rm{a}\gamma}\:|-1.19\,g_{\rm{aN}}^{0}+
  g_{\rm{aN}}^{3}|<1.36\times 10^{-16}$~GeV$^{-1}$ for $m_{\rm{a}}<0.03$~eV at
  the 95\% confidence level.

\vspace{3mm}                                
\begin{flushleft} \textbf{Keywords}: axions, solar physics
\end{flushleft}                             
\end{abstract}


\maketitle

\section{Introduction}                             \label{sec:intro}
Quantum chromodynamics (QCD), one of the most profound theories in modern physics and nowadays universally believed to be the theory of strong interactions, has one serious blemish: the ``strong CP problem". It arises from the fact that the QCD Lagrangian has a non-perturbative term (the so-called ``$\Theta$-term") which explicitly violates CP invariance in strong interactions. A very credible, and perhaps the most elegant solution to the strong CP problem was proposed by Peccei and Quinn in 1977 \cite{Pec77,Pec77a}. It is based on the hypothesis that the Standard Model has an additional global $U(1)$ chiral symmetry, now known as PQ (Peccei-Quinn) symmetry $U(1)_{\rm PQ}$, which is spontaneously broken at some large energy scale $f_{\mathrm a}$. An inevitable consequence of the Peccei-Quinn solution is the existence of a new neutral pseudoscalar particle, named the axion, which is the Nambu-Goldstone boson of the broken $U(1)_{\rm PQ}$ symmetry~\cite{Weinberg77,Wil78}. Due to the $U(1)_{\rm PQ}$ symmetry not being exact at the quantum level, as a result of a chiral anomaly, the axion is not massless and is, more precisely, a pseudo Nambu-Goldstone boson.

The phenomenological properties of the axion are mainly determined by the  scale $f_{\mathrm a}$ and closely related to those of the neutral pion. The axion mass is given by the relation
\begin{equation}
   m_{\rm a} = \frac{\sqrt{z}}{1+z}\, \frac{f_{\pi} m_{\pi}}{f_{\rm a}}
        =  6\,{\rm eV}\, \left(\frac{ 10\,^6\,\rm GeV}{f_{\rm a}}\right)\,,
   \label{eq:ax_mass}
\end{equation}
where $z=0.56$ is assumed for the mass ratio of the up and down quarks, while $f_{\mathrm \pi}\simeq 92$~MeV and $m_{\mathrm \pi}=135$~MeV are the pion decay constant and  mass respectively. Furthermore, the effective axion couplings to ordinary particles (photons, nucleons, and electrons) are inversely proportional to $f_{\mathrm a}$ as well, but they also include significant uncertainties originating from some model-dependent numerical parameters. It was originally thought that the energy scale of the $U(1)_{\rm PQ}$ symmetry breaking is equal to the electroweak scale, i.e. $f_{\mathrm a} = f_{\mathrm {ew}}$, with $f_{\mathrm {ew}}\simeq$ 250 GeV. The existence of the axion corresponding to such a scale, known as the ``standard" axion, was soon excluded by a number of experiments. Despite the failure of the standard axion model, it was possible to retain the Peccei-Quinn idea by introducing new axion models which decouple the $U(1)_{\rm PQ}$ scale from the electroweak scale, assuming that $f_{\mathrm a}$ has an arbitrary value much greater than 250 GeV, extending up to the Planck scale of $10^{19}$ GeV. Since the axion mass and its coupling constants with matter and radiation all scale as $1/f_{\mathrm a}$, the axion in the models with $f_{\mathrm a}\gg f_{\mathrm {ew}}$ is very light, very weakly coupled, and very long-lived, which makes it extremely hard to detect directly. This is why such models are generically referred to as ``invisible" axion models and they are still viable. Two classes of invisible axion models are often discussed in the literature: KSVZ (Kim, Shifman, Vainshtein, and Zakharov) or hadronic axions~\cite{Kim79,Shi80} and DFSZ (Dine, Fischler, Srednicki, and Zhitnitski\u{\i}) or GUT axions~\cite{Din81,Zhi80}. The main difference between the KSVZ and DFSZ-type axions is that the former do not couple to ordinary quarks and leptons at the tree level. However, due to the axion's interaction with photons, there is a radiatively induced coupling to electrons present for this type of axions, which is a process of higher order and hence extremely weak. As far as the interaction with photons is concerned, once the scale $f_{\mathrm a}$ is fixed, the axion-photon coupling constant $g_{{\rm a}\gamma}$ for DFSZ axions is also fixed, while hadronic axion models suggest different values for $g_{{\rm a}\gamma}$. Consequently, this coupling can be either suppressed or enhanced.

Depending upon the assumed value for $f_{\mathrm a}$, the existence of axions could have interesting consequences in astrophysics and cosmology. The emission of axions produced in the stellar plasma via processes based on their couplings to photons, electrons, and nucleons would provide a novel energy-loss mechanism for stars. This could accelerate the evolutionary process of stars, and thereby shorten their lifetimes. Axions may also exist as primordial cosmic relics copiously produced in the very early Universe, which makes them interesting candidates for the non-baryonic dark matter.

 So far the axion has remained elusive after over 30 years of intensive research, and none of the direct laboratory searches has been able to yield a positive signature for the axion. However, data from numerous laboratory experiments and astrophysical observations, together with the cosmological requirement that the contribution to the mass density of the Universe by relic axions does not overclose the Universe, restrict the allowed values of axion mass to a rather narrow range of $10^{-5}\, {\mathrm {eV}} < m_{\mathrm a}< 10^{-2}\, {\mathrm {eV}}$, but with uncertainties on either side. Thus the question of whether axions really do exist or the Peccei-Quinn mechanism is not realized in Nature still remains open, and the exhaustive search for axions continues. Detailed and updated reviews of the axion theory and experiments can be found in~\cite{PDG08,Duffy09,Ziout09}.

It is expected that pseudoscalar particles like axions should be copiously produced in stars by nuclear reactions and thermal processes in the stellar interior. A powerful source of axions would be the Sun in particular. As the closest and the best known astrophysical object, it is the source of choice for axion searches. Axions or similar axion-like particles (ALPs) that couple to two photons could be produced in the core of the Sun via Primakoff conversion of thermal photons in the electric and magnetic fields of the solar plasma. Such axions would have a continuous energy spectrum peaked near the mean energy of 4.2 keV and dying off above $\sim$10 keV. Most of the experiments that have been designed to search for these axions are based on the coherent axion-to-photon reconversion in a laboratory transverse magnetic field (the axion helioscope method~\cite{Ziout09,Sikivie:1983ip,vanBibber:1988ge,Laz92,Mori98,Ino02,Ino08,Zio99,Zio05,Andri07,Arik09}), or in the intense Coulomb field of nuclei in a crystal lattice of the detector (the Bragg scattering technique~\cite{Paschos:1993yf,Avi98,Mor02,Ber01,Ahmed09}).

 Due to the axion-nucleon coupling, there is an additional component of solar axions. If some nuclei in the Sun are excited either thermally (e.g. those with low-lying levels like $^{57}$Fe and $^{83}$Kr) or as a result of nuclear reactions (e.g. $^{7}$Li$^{*}$ nuclei produced by the $^{7}$Be electron capture), axion emission during their nuclear de-excitations could be possible. Such axions would be monoenergetic since their energy corresponds to the energy of the particular nuclear transition which produced them. With axions being pseudoscalar particles, the allowed values of angular momentum and parity carried away by them in such nuclear transitions are $0^{-}, 1^{+}, 2^{-},\ldots,$ which means that axions can be emitted in magnetic nuclear transitions. Up to now, these monoenergetic solar axions have mostly been searched for by using the resonant axion absorption process in targets made of the same nuclides in a terrestrial laboratory~\cite{Krc98,Krc01,Jak04,Der05,Nam07,Der07,Belli08}, or via processes based on axion-electron interactions, like Compton conversion of axion to photon and the axioelectric effect~\cite{Lju04,Borex08}.

In this paper we present the results of our search for monoenergetic 14.4 keV solar hadronic axions by using the CERN Axion Solar Telescope (CAST) setup. We assumed that such particles could be emitted from the Sun by de-excitation of thermally excited $^{57}$Fe nuclei~\cite{Hax91}. This stable isotope of iron (natural abundance 2.2 \%) is expected to be a suitable emitter of solar axions. It is exceptionally abundant among heavy elements in the Sun (solar abundance by mass fraction $2.8 \times 10^{-5}$). Also, its first excited nuclear state ($E^{*}=14.4$ keV) is low enough to be relatively easily thermally excited in the hot interior of the Sun ($kT\sim 1.3$ keV). The excited $^{57}$Fe nucleus relaxes to its ground state mainly through the emission of the 14.4 keV photon or an internal conversion electron. Since this de-excitation occurs dominantly via an M1 transition (E2/M1 mixing ratio is $0.002$), an axion could also be emitted.

In our attempt to detect $^{57}$Fe solar axions we relied on the axion helioscope method by using the latest and currently the most sensitive solar magnetic telescope, CAST, which is located at CERN. When its 9.26 m long LHC dipole test magnet is oriented towards the Sun, solar axions could convert to photons of the same energy via inverse Primakoff process while traversing the 9 T magnetic field produced in the two parallel bores inside the magnet. We have searched for a 14.4 keV peak in a spectrum recorded by the X-ray detector placed at the far end of the magnet facing the apertures of the bores. If observed, this peak could be interpreted as the result of the conversion of the $^{57}$Fe solar axions into photons inside the magnet bores, and hence as the direct signal for such axions. As a contrast to the previous searches for these axions, which relied solely on the axion-nucleon coupling, our search involved not only the axion-nucleon interaction (in the emission process) but also the axion-photon interaction (in the detection process). This allowed us to explore the relation between axion-photon and axion-nucleon coupling constants for a wide range of axion masses. Although we focus on the axion because of its  theoretical motivation, our results also apply to similar pseudoscalar particles that couple to two photons and can be emitted in the nuclear magnetic transition.

\section{Axion emission from $^{57}$Fe nuclei in the Sun}                          \label{sec:axion_emission}

The axion-nucleon coupling arises from two contributions : the tree-level coupling of the axion to up- and down-quarks, and a contribution due to the generic axion-pion mixing, a phenomenon which is the result of the axion-gluon coupling, and the fact that axion and pion are bosons with the same quantum numbers, so they mix. This means that even a hadronic axion has a coupling to nucleon, although it does not couple directly to ordinary quarks.\\
The effective Lagrangian for the axion-nucleon interaction can be written as
\begin{eqnarray} \label{eq:axion_nucleon_L}
{\cal{L}}_{\rm{aN}}=i\,a\,\bar{\psi}_{\rm{N}}\gamma_{5}
(g_{\rm{aN}}^{0}+g_{\rm{aN}}^{3}\tau_{3}) \psi_{\rm{N}}\,,
\end{eqnarray}
where $a$ is the axion field, $\psi_{\rm{N}}=${\scriptsize$\left( \begin{array}{c}
p \\
n \end{array}\right)$} is the nucleon doublet, and $\tau_{3}$ is the Pauli matrix. The isoscalar $g_{\rm{aN}}^{0}$ and isovector $g_{\rm{aN}}^{3}$ axion-nucleon coupling constants are model dependent, i.e., they depend on the details of the theory implementing the Peccei-Quinn mechanism. For example, in hadronic axion models they are related to the scale $f_{\mathrm a}$ by expressions~\cite{Kap85,Sred85}
\begin{eqnarray} \label{eq:ax_nucl_0}
g_{\rm{aN}}^{0}= -\frac{m_{\rm{N}}}{f_{\rm{a}}} \frac{1}{6}\, \left( 2S + 3F - D\right)
\end{eqnarray}
and
\begin{eqnarray} \label{eq:ax_nucl_3}
g_{\rm{aN}}^{3}=-\frac{m_{\rm{N}}}{f_{\rm{a}}} \frac{1}{2}\,(F+D)\frac{1-z}{1+z}\,.
\end{eqnarray}
Here, $m_{\rm{N}}$ is the nucleon mass, while the constants $F=0.462$ and $D=0.808$~\cite{Mateu05} are the antisymmetric and symmetric reduced matrix elements for the $SU(3)$ octet axial-vector currents. They are determined from the hyperon semileptonic decays and flavor $SU(3)$ symmetry. The flavor-singlet axial-vector matrix element $S$ is still a poorly constrained parameter. It can be estimated from the polarized nucleon structure functions data, but suffers from large uncertainties and ambiguity.

We focused our attention on the decay of the 14.4 keV first excited state of $^{57}$Fe nucleus to the ground state via axion emission, a process that competes with ordinary M1 and E2 gamma decay. In general, the axion-to-photon emission rate ratio for the M1 nuclear transition calculated in the long-wavelength limit is~\cite{Avignone88}
\begin{eqnarray} \label{eq:axionphotonration}
\frac{\Gamma_{\rm{a}}}{\Gamma_{\gamma}} = \left(
\frac{k_{\rm{a}}}{k_{\gamma}}\right) ^{3} \: \frac{1}{2\pi\alpha}\: \frac{1}{1+\delta^{2}}\: \left[\frac{g_{\rm{aN}}^{0}\beta +
g_{\rm{aN}}^{3}}{(\mu_{0}-1/2)\beta + \mu_{3} - \eta}\right] ^{2}\,,
\end{eqnarray}
where $k_{\rm{a}}$ is the momentum of the outgoing axion, $k_{\gamma}$ represents the photon momentum, and $\alpha\simeq 1/137$ is the fine structure constant. The quantities $\mu_{0}$=0.88 and $\mu_{3}$=4.71 are the isoscalar and isovector nuclear magnetic moments respectively, given in nuclear magnetons. The parameter $\delta$ denotes the E2/M1 mixing ratio for the particular nuclear transition, while $\beta$ and $\eta$ are nuclear structure dependent ratios. Their values for the 14.4~keV de-excitation process of an $^{57}$Fe nucleus are $\delta$=0.002, $\beta=-1.19$, and $\eta=0.8$~\cite{Hax91}. Using these values in equation~(\ref{eq:axionphotonration}) we find
\begin{eqnarray} \label{eq:axionphotonratio2}
\frac{\Gamma_{\rm{a}}}{\Gamma_{\gamma}} = 1.82\;(-1.19g_{\rm{aN}}^{0}+g_{\rm{aN}}^{3})^{2}\,.
\end{eqnarray}
 In the above expression we made approximation $(k_{\rm{a}}$/$k_{\gamma})^{3}\simeq 1$ for the phase space factor ratio. This is based on the assumption that the axion mass is negligible compared to the axion energy which equals, in our case, the transition energy of 14.4 keV. Since the coupling constants $g_{\rm{aN}}^{0}$ and $g_{\rm{aN}}^{3}$  are model dependent, we can consider the parameter $g_{\rm{aN}}^{\rm{eff}}\equiv(-1.19g_{\rm{aN}}^{0}+g_{\rm{aN}}^{3})$ as a free unknown parameter characterizing not only the axion-nucleon coupling but also, more generally, the nucleon coupling to any axion-like particles that could be emitted in the M1 nuclear transition. In terms of the axion mass, equation~(\ref{eq:axionphotonratio2}) can be expressed by combining equations~(\ref{eq:ax_mass}), (\ref{eq:ax_nucl_0}) and (\ref{eq:ax_nucl_3}) as
\begin{eqnarray} \label{eq:axionphotonratio3}
\frac{\Gamma_{\rm{a}}}{\Gamma_{\gamma}} = 3.96\times 10^{-16}\,\left(\frac{m_{\rm{a}}}{1\,\mathrm{eV}}\right)^{2}\,,
\end{eqnarray}
where $S=0.4$~\cite{Mallot00} is assumed.

To estimate the flux of monoenergetic solar axions emitted from $^{57}$Fe nuclear de-excitations, the calculation was performed as in \cite{Hax91,Mor95}.
 The $^{57}$Fe nucleus can be thermally excited to its first excited state in the Sun since its excitation energy, $E^{*}$=14.4~keV, is comparable to the Sun's core temperature ($kT\sim 1.3$ keV). The probability for this thermal excitation is given by the Boltzmann distribution
\begin{eqnarray}  \label{eq:Boltzmann}
w_{1}\simeq\frac{(2J_{1}+1)\,e^{-E^{*}\!\!/kT}}{(2J_{0}+1)+(2J_{1}+1)\,e^{-E^{*}\!\!/kT}}\,,
\end{eqnarray}
where $J_{1}=3/2$ and $J_{0}=1/2$ are total angular momenta of the first excited and ground states respectively. In further calculations we can neglect the second term in the denominator because $e^{-E^{*}\!\!/kT}\ll 1$.
The 14.4 keV axion emission rate per unit mass due to the de-excitation of thermally excited $^{57}$Fe nuclei in the Sun is
\begin{eqnarray} \label{eq:axionrate}
{\cal{N}}_{\rm{a}}={\cal{N}}\;w_{1}\; \frac{1}{\tau_{\gamma}}\;\frac{\Gamma_{\rm{a}}}{\Gamma_{\gamma}}\;\;\;\;\;\mathrm{g^{-1}\;s^{-1}}\,,
\end{eqnarray}
where ${\cal{N}}=3.0\times 10^{17}$ g$^{-1}$ is the number of $^{57}$Fe nuclei per 1 g of solar matter~\cite{Turck93}, and $\tau_{\gamma}=1.3\times 10^{-6}$ s is the mean lifetime of the first excited nuclear state of $^{57}$Fe associated with the partial gamma decay width of that state. Owing to Doppler broadening caused by the thermal motion of $^{57}$Fe nuclei in the hot solar interior, the axion emission spectrum is a Gaussian with the standard deviation parameter
\begin{eqnarray} \label{eq:doppler}
\sigma(T)=E^{*}\sqrt{\frac{kT}{m}}\,,
\end{eqnarray}
where $T$ denotes the temperature at the location in the Sun where the axion is produced, while $m$ is the mass of the $^{57}$Fe nucleus. This spectrum is approximately  centered at the transition energy $E_{1}=E^{*}=14.4$ keV because the axion energy shift, caused by the recoil of the $^{57}$Fe nucleus in the emission process ($\sim 1.9 \times 10^{-3}$ eV) and the gravitational redshift due to the Sun ($\sim 1.5\times 10^{-1}$ eV), is negligible compared with the Doppler broadening of the spectrum (FWHM$=2.35\; \sigma(T)\sim 5$ eV). Following these arguments, we can write the differential $^{57}$Fe solar axion flux expected at the Earth as
\begin{eqnarray} \label{eq:diff_flux}
\frac{d\Phi_{\rm{a}}(E_{\rm{a}})}{dE_{\rm{a}}}&=&\frac{1}{4\pi d_{\odot}^{\,2}}\,
\int_{0}^{R_{\odot}}{\cal{N}}_{\rm{a}}\, \frac{1}{\sqrt{2\pi}\sigma(T(r))}\,\mathrm{exp}\left[-\frac{(E_{\rm{a}}-E_{1})^{2}}{2\sigma(T(r))^{2}}\right] \times \nonumber\\
& & \times  \rho(r)\, 4\pi r^{2}\, dr\,.
\end{eqnarray}
Here $E_{\rm{a}}$ is the axion energy, $d_{\odot}$ denotes the average distance between the Sun and the Earth, $R_{\odot}$ is the solar radius,
while $T(r)$ and $\rho(r)$ are the temperature and the mass density in a spherical shell with the radius $r$ in the solar interior, respectively.

The total flux of $^{57}$Fe solar axions at the Earth can be calculated by integrating the differential axion flux  (equation (\ref{eq:diff_flux})) with respect to the axion energy. This leads to
\begin{eqnarray}
\Phi_{\rm{a}}&=&\int \frac{d\Phi_{\rm{a}}(E_{\rm{a}})}{dE_{\rm{a}}}\, d E_{\rm{a}} \nonumber \\
&=&\frac{1}{4\pi d_{\odot}^{\,2}}\,
{\cal{N}}\, \frac{1}{\tau_{\gamma}}\,
\frac{\Gamma_{\rm{a}}}{\Gamma_{\gamma}}\,\, 2\,
\int_{0}^{R_{\odot}}e^{-E^{*}\!/kT(r)}\, \rho(r)\, 4\pi r^{2}\: dr\,,
\end{eqnarray}
with the help of equations (\ref{eq:Boltzmann}) and (\ref{eq:axionrate}). As in the framework of the standard solar model, we assumed that the number of $^{57}$Fe nuclei per unit mass in the Sun is uniform, i.e., that ${\cal{N}}$ is independent of $r$. Using the standard solar model data for the temperature and mass density distributions in the Sun as functions of the radius $r$~\cite{SSM}, we evaluated the integral in the above expression and found that
\begin{eqnarray} \label{eq:total_flux}
\Phi_{\rm{a}}=4.56\times 10^{23}\,(g_{\rm{aN}}^{\rm{eff}})^{2}\;\;\;\;\;\mathrm{cm}^{-2}\;\mathrm{s}^{-1}\,.
\end{eqnarray}
The corresponding solar axion luminosity is calculated to be \footnote{Following the calculations in~\cite{Raff86} we estimated that processes like the Compton process on nuclei and proton-Helium bremsstrahlung, that are also based on the axion-nucleon coupling, are suppressed relative to the hadronic axion emission from the M1 nuclear transition of $^{57}$Fe.}
\begin{eqnarray}  \label{eq:ax_luminosity}
L_{\rm{a}}=7.68\times 10^{9}(g_{\rm{aN}}^{\rm{eff}})^{2}\, L_{\odot}\,,
\end{eqnarray}
where $L_{\odot}=3.84\times 10^{26}$~W is the solar photon luminosity. Arguments related to the measured solar neutrino flux ($L_{\rm a}<0.1\,L_{\odot}$)~\cite{Gondolo09} lead to
\begin{eqnarray}  \label{eq:constrain}
\left|g_{\rm{aN}}^{\rm{eff}}\right|<3.6\times 10^{-6}\,.
\end{eqnarray}

\section{The detection of solar axions in the CAST experiment}                          \label{sec:cast_experiment}
The CERN Axion Solar Telescope (CAST) experiment is the most recent  implementation of the axion helioscope technique~\cite{Sikivie:1983ip}. It searches for solar axions and similar particles with unprecedented sensitivity in the sub-eV mass range which is comparable to the astrophysical constraints on these particles. Like most of the axion search experiments in the past 30 years, this experiment relies on the axion coupling to two photons, a generic property of axions and ALPs given by the effective Lagrangian
\begin{eqnarray}  \label{eq:axion_photon_lagrangian}
{\cal{L}}_{\rm{a\gamma}}=-\frac{1}{4}\, g_{\rm{a\gamma}}\,F^{\mu\nu}\,\tilde{F}_{\mu\nu}\,a=g_{\rm{a\gamma}}\, {\bf{E}}\cdot{\bf{B}}\,a\,,
\end{eqnarray}
where $a$ is the axion field, $F^{\mu\nu}$ the electromagnetic field strength tensor, and $\tilde{F}_{\mu\nu}$ its dual. This interaction can also be expressed in terms of electric $\bf{E}$ and magnetic $\bf{B}$ field of the coupling photons, as shown in the above expression. The effective axion-photon coupling constant $g_{\rm{a\gamma}}$ is given by
\begin{eqnarray} \label{eq:gagamma}
g_{\rm{a\gamma}}=\frac{\alpha}{2\pi f_{\rm{a}}} \left[ \frac{E}{N}-\frac{2\,(4+z)}{3\,(1+z)}\right] =
\frac{\alpha}{2\pi f_{\rm{a}}} \left( \frac{E}{N}-1.95\pm 0.08\right) \,.
\end{eqnarray}
Here $E$ and $N$ are the model dependent coefficients of the electromagnetic and color anomaly of the axial current associated with the $U(1)_{\rm PQ}$ symmetry, respectively. In DFSZ axion models their ratio is fixed to $E/N=8/3$, while for the hadronic axions this ratio can take different values depending on the details of each model.

Axions and similar particles with a two-photon interaction of the form given by equation (\ref{eq:axion_photon_lagrangian}) can transform into photons, and vice versa, in external electric or magnetic fields. Therefore, these particles could be produced in stars by the Primakoff conversion of thermal photons in the Coulomb fields of nuclei and electrons in the stellar plasma. On the other hand, the reverse process has served as the basis for various experimental methods to search for these particles. The axion helioscope technique~\cite{Sikivie:1983ip} is one such method. The essence of this idea is to search for solar axions using a long dipole magnet in a laboratory. Inside the magnet, while it is oriented towards the Sun, the incoming axion couples to a virtual photon provided by the transverse magnetic field and converts into a real photon via the reverse Primakoff process  $a+\gamma_{\mathrm{virtual}}\rightarrow\gamma$. This photon has the energy equal to the axion energy and can be detected with a suitable X-ray detector placed at the far end of the magnet opposite the Sun. Assuming that there is a vacuum in the conversion volume (i.e. magnetic field region), the axion-photon conversion probability is~\cite{Laz92}\footnote{We use natural units with $\hbar=c=1$.}
\begin{eqnarray} \label{eq:conversion_prob}
P_{\rm{a\gamma}}=\left( \frac{g_{\rm{a\gamma}}\,B\,L}{2}\right)^{2}
  \frac{4}{q^{2}\:L^{2}}\; \sin^{2}\left( \frac{q\:L}{2}\right)\,,
\end{eqnarray}
where $B$ is the magnetic field, $L$ is the length of the conversion volume in the direction of the incoming axion propagation, and $q=m_{\rm{a}}^{2}/2E_{\rm{a}}$ represents the momentum difference between the axion and the photon of energy $E_{\rm{a}}$. The conversion probability is maximal if the axion and photon fields remain in phase over the length of the conversion region, i.e., when the coherence condition $qL<\pi$ is satisfied~\cite{Laz92}. This restricts the sensitivity of a helioscope to a specific range of axion masses, e.g., for a 10 m long magnet and axion energy of $\sim$10 keV the coherence condition sets the limit of $m_{\rm{a}}\lesssim 0.03$\nolinebreak[4] eV on the axion mass, up to which such an experiment is sensitive. However, coherence can be maintained for higher axion masses if the conversion volume is filled with a buffer gas such as helium~\cite{vanBibber:1988ge}. In this case, photons acquire an effective mass $m_{\gamma}$ whose value is determined by the gas pressure, and the axion-photon momentum difference becomes $q=\left| m_{\gamma}^{2}-m_{\rm{a}}^{2}\right|/2E_{\rm{a}}$. As a result, the coherence is restored for a narrow axion mass range $m_{\rm{a}}\simeq m_{\gamma}$, where the effective photon mass matches the axion mass.

The first experiment to use the axion helioscope technique was performed at BNL in the early 1990s~\cite{Laz92}. Following this experiment, the Tokyo Axion Helioscope continued the search for axions using the same method with much improved sensitivity~\cite{Mori98,Ino02,Ino08}. At present, the most sensitive axion helioscope is the CERN Axion Solar Telescope (CAST). The main component of CAST is a decommissioned prototype of a twin aperture LHC superconducting dipole magnet, which serves as a magnetic telescope. It provides a transverse magnetic field of 9.0 T inside the two parallel, straight, 9.26 m long bores. The aperture of each bore is 43 mm, so the total cross-sectional area is $2\times 14.5$ cm$^{2}$. In terms of the parameter $(BL)^{2}$, which according to equation (\ref{eq:conversion_prob}) determines the axion-photon conversion probability, the CAST magnet is  $\sim\,$80 times more efficient as an axion-to-photon converter than the best competing helioscope in Tokyo. To optimize CAST's performance, the magnet is installed on a moving platform which allows it to track the Sun $\pm 8^{\circ}$ vertically and $\pm 40^{\circ}$ horizontally. Thus, it can be aligned with the Sun for approximately 1.5 h both during sunrise and sunset every day throughout the year. In order to detect photons coming from the magnet bores, as a result of axion conversion in the magnetic field, several low-background X-ray detectors are installed on both ends of the magnet. Until 2007, a conventional Time Projection Chamber (TPC) was located at one end, covering both magnet bores, to detect photons originating from axions during the tracking of the Sun at sunset. It was then replaced by two MICROMEGAS detectors, each attached to one bore. On the other side of the magnet, there is another MICROMEGAS detector covering one bore, and an X-ray mirror telescope with a pn-CCD chip as the focal plane detector at the other bore, both intended to detect photons produced from axions during the sunrise solar tracking. More details about the CAST experiment and detectors can be found in~\cite{Zio99,Zio05,Andri07,Arik09,Kus07,Aut07,Abb07}.

To cover as wide as possible range of potential axion masses, the operation of the CAST experiment is divided into two phases. During the Phase I (2003--2004)~\cite{Zio05,Andri07} the experiment operated with vacuum inside the magnet bores and the sensitivity was essentially limited to $m_{\rm{a}}<0.02$ eV due to the coherence condition. In the second phase (so-called Phase II) which started in 2005, the magnet bores are filled with a buffer gas in order to extend the sensitivity to higher axion masses. In the first part of this phase (2005--2006) $^{4}$He was used as a buffer gas. By increasing the gas pressure in appropriate steps, axion masses up to $\sim$0.4 eV were scanned and the results of these measurements supersede all previous experimental limits on the axion-photon coupling constant in this mass range~\cite{Arik09}. To explore axion masses above 0.4\nolinebreak[4] eV, $^{3}$He has to be used because it has a higher vapor pressure than $^{4}$He. This allows us to further increase gas pressure in the magnet bores and to reach axion masses up to about 1 eV in the ongoing second part of Phase II that started in 2007 and is planned to finish by the end of 2010.

\section{Measurement and data analysis}                          \label{sec:measurements_data_analysis}
The CAST experiment is primarily designed to search for axions or axion-like particles that could be produced in the Sun by the Primakoff conversion of thermal photons. Their expected energy spectrum~\cite{Andri07},
\begin{equation} \label{eq:Prim_flux}
\frac{d\Phi_{\rm{a}}^{\rm{P}}(E_{\rm{a}})}{dE_{\rm{a}}}=6.02\times 10^{30}\,g_{{\rm a}\gamma}^2\,E_{\rm a}^{2.481}\,e^{-E_{\rm a}/1.205} \;\;\;\;{\rm cm}^{-2}\;{\rm s}^{-1}\;{\rm keV}^{-1}\,,
\end{equation}
(where energies are in keV) has the peak at 3~keV, mean energy of 4.2~keV, and vanishes above 10~keV. Since the conversion of these particles inside the CAST magnet bores would produce photons of the same energies, the X-ray detectors used in CAST are optimized for the efficient detection of photons in the 1-10 keV range. To search for 14.4 keV photons that might originate from the conversion of the $^{57}$Fe solar axions we used only the data provided by the TPC detector because the other detectors have very low sensitivity to photons with energies above $\sim$10 keV.

The CAST TPC incorporates the well-known concepts of drift chambers and Multi-Wire Proportional Chambers (MWPC). It has an active volume of $10\times 15\times 30$ cm$^{3}$ filled with an Ar (95\%) + CH$_{4}$ (5\%) gas mixture at atmospheric pressure, where the incoming particle interacts with the gas producing free electrons. The detector's volume is 10 cm long in the drift direction that is parallel to the magnet axis, while the $15\times 30$ cm$^{2}$ cross section, covering both magnet bores, is perpendicular to this direction. The drift region is bounded on the front side with a $15\times 30$ cm$^{2}$ drift electrode, biased at -7 kV, that is made of a thin aluminum layer and covers the entire inner side of the chamber wall closest to the magnet. On the opposite end, i.e., back side of the chamber, there are 3 planes of sense wires parallel to the drift electrode: one anode plane at +1.8\nolinebreak[4] kV with 48 wires placed between two grounded cathode planes with 96 wires each. The wires in both cathode planes are perpendicular to the anode wires. The spacing between two adjacent wires of the same plane is 3 mm. The gap between the anode and the inner cathode plane (the one closest to the drift electrode) is 3 mm, while the distance between the anode and the outer cathode plane is 6 mm. This asymmetric configuration enhances the induced signals on the cathode wires in the inner plane, which are the ones being read out by the front-end electronics, together with the anode wires. Each of these wires is read out individually, and since the anode wires are perpendicular to the cathode wires, a very good two-dimensional position resolution can be obtained. This allows us to distinguish spatially localized photon events that may represent the axion signal from the background events characterized by the long tracks of cosmic rays. The detector is connected to the magnet by two 6 cm diameter entrance windows made of a thin aluminized mylar foil stretched on a metallic grid. These windows allow the photons coming from the magnet bores to enter the detector. For a detailed description of the TPC detector, its shielding, front-end electronics, and data acquisition system, we refer to~\cite{Aut07,Luz07}.

\begin{figure*}[!ht]
  \begin{minipage}{0.49\textwidth}
    \centerline{\includegraphics[width=0.95\textwidth]{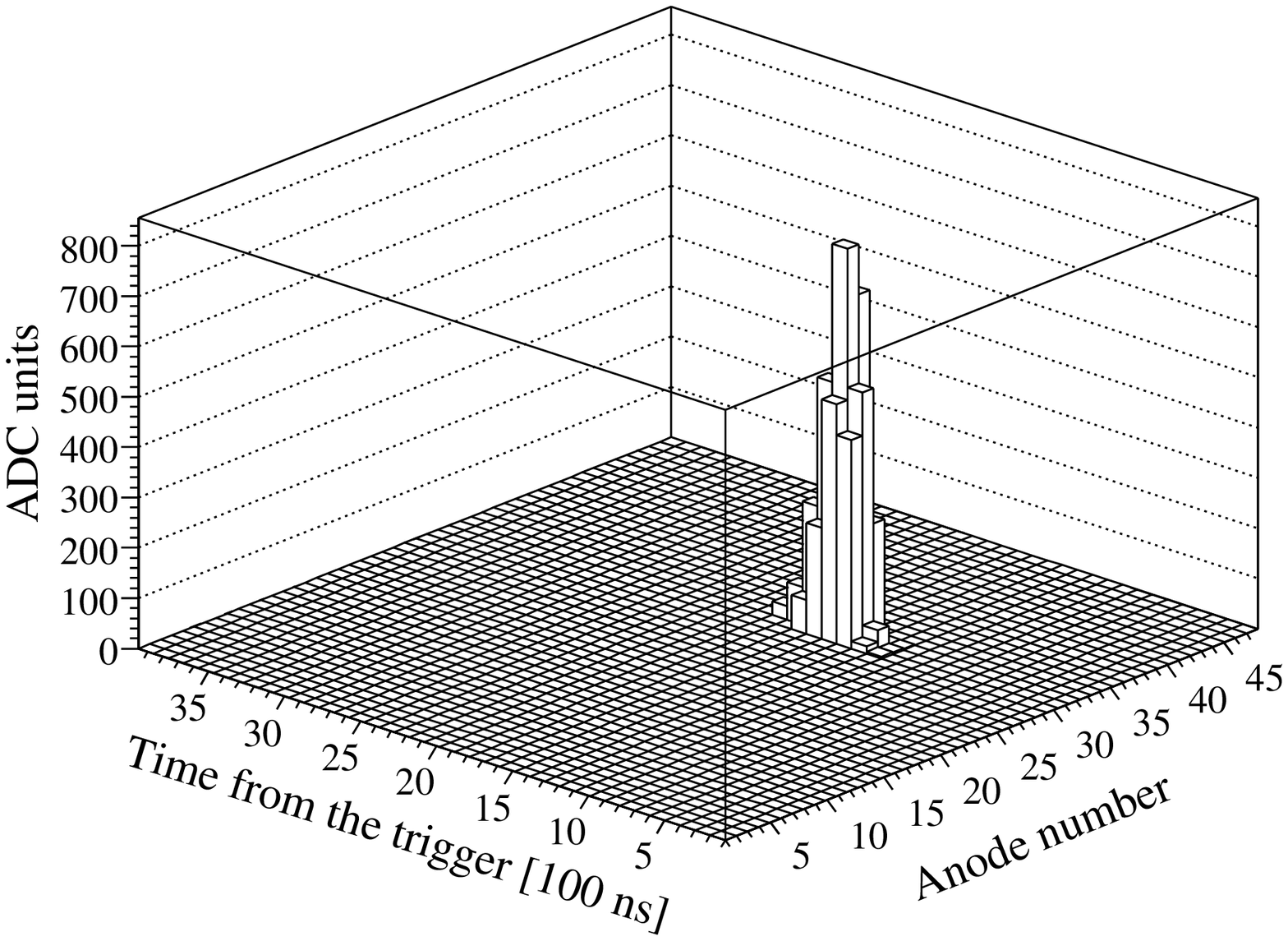}}
  \end{minipage}
  \begin{minipage}{0.49\textwidth}
    \centerline{\includegraphics[width=0.95\textwidth]{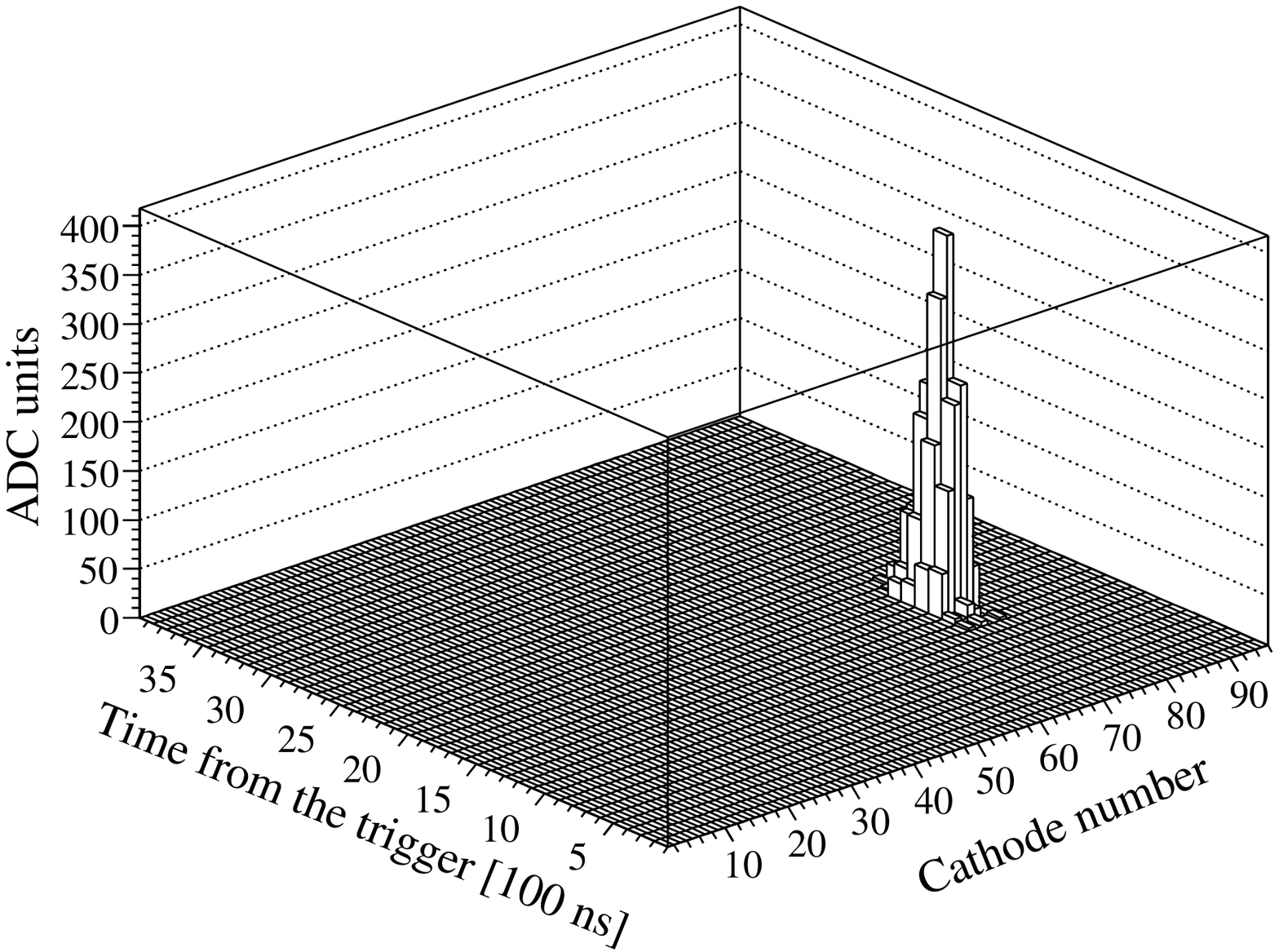}}
  \end{minipage}
 \caption{ Time evolution of the pulse induced in each wire as recorded by the flash-ADCs in a photon-like event. Left (right) plot shows pulses in anode (cathode) wires.
 \label{fig:signal_profile}}
\end{figure*}

To characterize the CAST TPC detector, a series of calibration measurements was performed using the X-ray beams with very accurately calibrated energies and intensities at the PANTER X-ray facility of MPE in Munich~\cite{Aut07}.  Photons originating from the axion conversion in the CAST magnet would enter the detector only through its two entrance windows, their direction being parallel to the magnet axis, i.e., perpendicular to the anode and cathode wire planes. Such a photon would normally produce a point-like energy deposition in the region of the detector's sensitive volume facing the windows. From the analysis of the PANTER data it was realized that the pulses induced on the sense wires (hits) in such event are clustered on  several contiguous anode wires (so\nolinebreak[4]-\nolinebreak[4]called anode signal cluster), as well as on several contiguous cathode wires (cathode cluster).
This characteristic profile of a photon event can be seen in the example shown in figure~\ref{fig:signal_profile}. The left and the right plot display the time evolution of the pulse induced in each anode and each cathode wire, respectively, as recorded by the flash-ADCs.
By studying the cluster properties and topology, and exploiting the two-dimensional position reconstruction capability of the detector, we were able to establish a set of rules (cuts) to distinguish genuine photon events that might be an axion signal from background events coming from cosmic rays and natural radioacti\-vity. These cuts take into consideration the total number of anode and cathode clusters in the event, their time correlation, cluster multiplicity given by the number of hits in each cluster, and the position of clusters in the anode and catho\-de wire planes with respect to the entrance windows of the detector~\cite{Aut07}. We should emphasize that these cuts were derived from the data obtained during calibration measurements at PANTER where monochromatic photon beams with energies from 0.3 to 8 keV were used. This energy range was chosen because it covers the bulk of the expected spectrum of solar axions produced by the Primakoff process, and these axions are, as stated before, the main subject of CAST's search for solar axions. It is expected that the similar cuts can be applied to single out 14.4 keV photon events that could indicate the conversion of $^{57}$Fe solar axions inside the bores of the CAST magnet. In order to verify this assumption, we made additional calibration measurements with a $^{57}$Co source placed in front of each of the two TPC's windows. $^{57}$Co was the source of choice for this task because in the course of the $^{57}$Co decay to the ground state of $^{57}$Fe, several photons are emitted, and one of them has the energy of 14.4 keV.

\begin{figure*}[t]
\begin{center}
 \centerline{\includegraphics[width=0.65\textwidth]{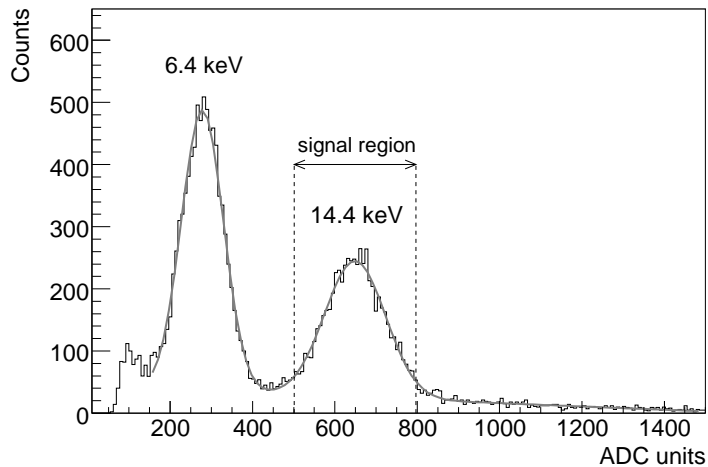}}
 \caption{ TPC calibration spectrum measured with $^{57}$Co source after the off-line analysis cuts were applied to the data. Two peaks are present: Fe X-rays (K$_{\mathrm{\alpha}}$) at 6.4~keV and 14.4 keV photons corresponding to the $\gamma$-transition in $^{57}$Fe. The grey line shows the combined fit of the peaks plus background while arrows indicate the region of interest for the axion signal.
 \label{fig:57Co_spectrum}}
 \end{center}
\end{figure*}

Figure~\ref{fig:57Co_spectrum} shows the spectrum of $^{57}$Co source after the selection cuts were applied to the TPC data. As a starting point to reject events that were not produced by photon interactions in the detector's volume, we followed the information obtained from the PANTER calibration measurements and set the requirement that only events with one anode cluster and one cathode cluster should be considered as relevant photon events. This condition also allowed us to straightforwardly match these two clusters in order to obtain the two-dimensional position of the point-like event. Our study of the cluster properties in the events that constitute the 14.4 keV peak in the $^{57}$Co calibration spectrum resulted in a set of additional cuts given in table ~\ref{tab:cuts}. The first two cuts consider the spread of the clusters. The spread of the anode cluster is due to the diffusion of the electron cloud along the drift distance from the interaction point of an incoming photon to the anode wires plane. The larger spread of the cathode cluster is due to the development of the avalanche process along the anode wires. The ``time correlation" cut reflects the fact that in a real photon event both anode and cathode cluster originate from the same avalanche process induced in the proximity of the anode wires by the initial ionization electron cloud. The purpose of this cut is to reject events with spurious clusters that might be produced by the effects of noise in contiguous wires and mimic the real photon events. The ``no saturation" cut is related to the fact that energy deposited in each event is calculated using the total strength of the cathode cluster obtained by adding up the pulse height of every hit in the cluster. If any of the hits has a pulse higher than the one that the flash-ADC can handle, the calculated energy would be incorrect, and thus such event should be rejected. Finally, the ``position"  cut is used to reject events with two-dimensional positions outside the region where the axion signal is expected, i.e., out of the area facing the two magnet bores.
The application of all these cuts in the off-line analysis of data recorded during normal CAST operation reduces the background by approximately two orders of magnitude with respect to the raw trigger rate without significantly reducing the efficiency of the detector.
\begin{table}[t]
  \centering
  \footnotesize
  \caption { List of software cuts applied to the CAST TPC
      data in the search for $^{57}$Fe solar axions.}
  \begin{tabular}{cc} \\
      \hline
      \bf{Cut} & \bf{Condition} \\ \hline \\
      Anode cluster multiplicity &  2 or 3 hits in the anode cluster \\
      Cathode cluster multiplicity & 2 to 8 hits in the cathode cluster \\
      Anode--cathode cluster time correlation & Time difference between anode and cathode cluster \\
      & must be in the range $-0.15$ to $0.02$ $\mu\text{s}$\\
      No saturation & No hit in the cathode cluster reaching the upper limit \\ & of the flash-ADC dynamical range\\
      Position & 2-D coordinates of the event should be inside \\ & the
      area facing the magnet bores \\ \hline
    \end{tabular}
    \label{tab:cuts}
\end{table}

The hardware efficiency of the TPC to detect 14.4~keV photons was estimated by Monte Carlo simulations, using the GEANT4 software package. The geometry of the detector, the transparency of the entrance windows, as well as the opacity of the gas inside the sensitive detector volume were taken into account, giving the efficiency of~13\%. Here we note that the validity of the simulations was checked by comparing the computed efficiencies with those obtained from the PANTER measurements~\cite{Aut07} for photon energies of 6.4 and 8~keV (the two highest energies at which measurements were performed). The agreement between the corresponding efficiencies is within 4 and 2 percent respectively. Therefore, we regarded the efficiency computed by simulation as a good approximation for the TPC hardware efficiency at 14.4 keV. The additional efficiency loss of 18\% in the off-\nolinebreak[4]line analysis, due to the software cuts applied to reduce the background, was calculated using $^{57}$Co calibration data. Hence, the overall (hardware + software) TPC detection efficiency of 10.7\% for 14.4~keV photons that may come from conversion of solar axions was obtained. Finally, data from $^{57}$Co calibration runs were also used to determine the energy resolution of the TPC at 14.4~keV which defines the energy region of interest in our search for the signal of $^{57}$Fe solar axions. It was found that the width of the 14.4~keV peak is characterized by the standard deviation parameter $\sigma=1.77$~keV, giving a full width at half maximum of 4.16~keV.

For the study presented in this article, we used the data acquired with the TPC over the period of $\sim$5.5 months in 2004 during Phase I of the CAST experiment.
The energy response of the detector was calibrated periodically using the $^{55}$Fe X-ray source. No systematic shift in the energy scale with time was found. Since the calibration runs took place every six hours, we were able to characterize, with very good precision, small gain variations due to the fluctuations of environmental parameters that affected the gain of the detector, and to correct the calculation of the energy deposited in each event accordingly.

Data quality checks were performed both on-line (during data acquisition) and off-line by using the ``slow control" data that were continuously recorded during the operation of the experiment to monitor various experimental parameters (e.g. the magnetic field strength, magnet pointing direction with respect to the current position of the Sun, magnet temperature, various pressures, etc.). As a result, data qualified for the analysis were accumulated during 2819 effective hours, out of which 203 hours correspond to the tracking data (data acquired in the ``axion-sensitive" conditions, i.e., while the magnet was tracking the Sun), and 2616 hours correspond to the background data (data acquired during the non-tracking periods). The background data were used to estimate and subtract the true background contribution in the tracking data spectrum. Therefore, these data were taken while the magnet was parked in well-defined positions close to where it was passing during the Sun tracking, but at times when the Sun was not in view.

\begin{figure*}[!ht]
  \begin{minipage}{0.49\textwidth}
    \centerline{\includegraphics[width=0.95\textwidth]{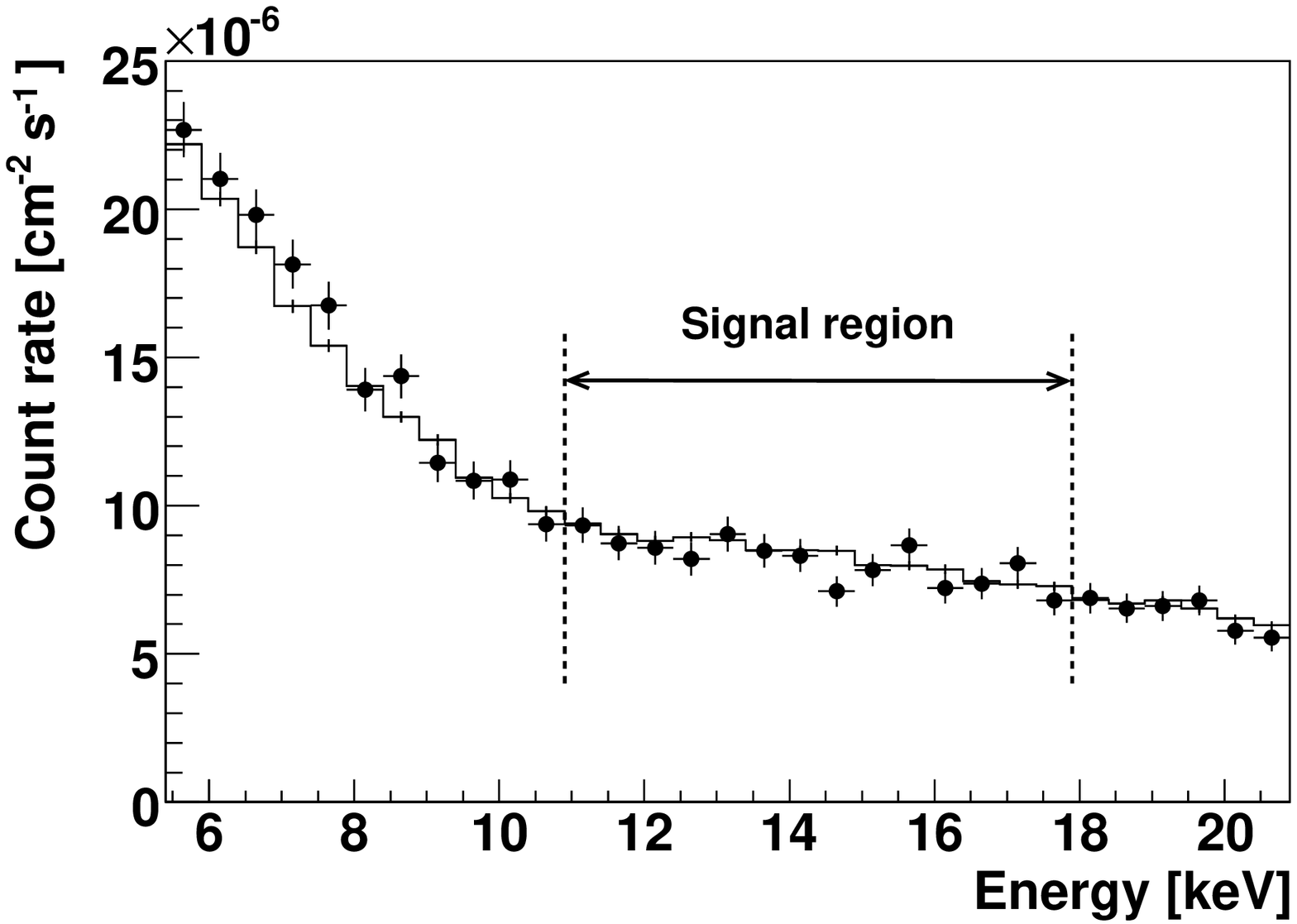}}
  \end{minipage}
  \begin{minipage}{0.49\textwidth}
    \centerline{\includegraphics[width=0.95\textwidth]{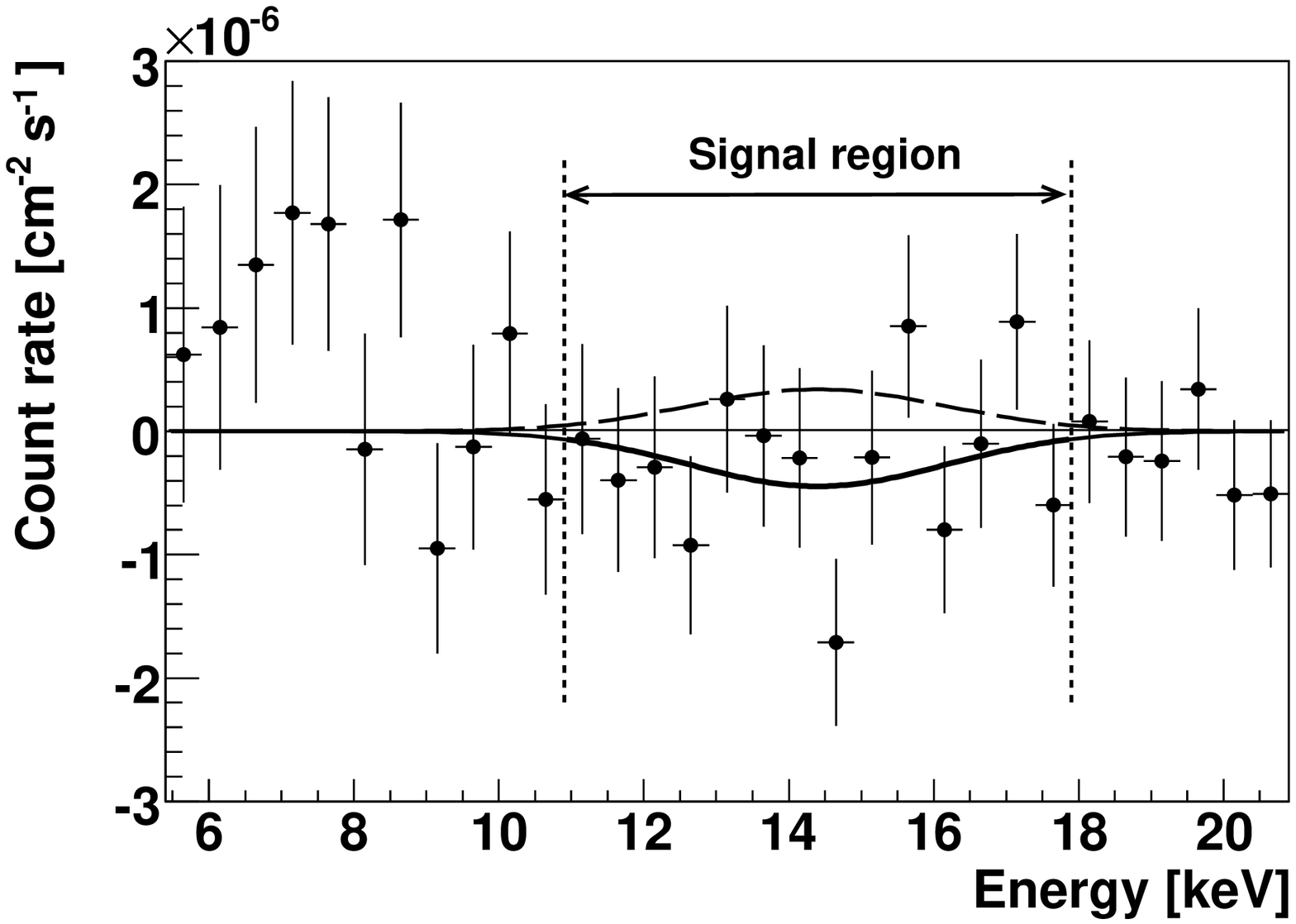}}
  \end{minipage}
 \caption{
   Left: Energy distribution of events recorded during 203 hours of Sun tracking ($\fullcircle$), i.e. while the TPC was sensitive to solar axion signals, compared to the background spectrum ($\full$) measured during non-tracking periods in total of 2616 hours. Right: The subtracted spectrum ($\fullcircle$) together with the expectation for the best fit $N_{\rm S}$ ($\full$) and for the 95\% C.L. limit on $N_{\rm S}$ ($\dashed$). The signal region ($14.4\pm 3.5$~keV) that is expected to cover the 14.4 keV peak containing photon events coming from the conversion of $^{57}$Fe solar axions in the magnet bores is also shown in both plots.
 \label{fig:spectra}}
\end{figure*}

The energy spectrum of the events reconstructed from the tracking data that passed all the software cuts in the off-line analysis, as well as the measured background spectrum after applying the same cuts, are shown in figure~\ref{fig:spectra} on the left side. Both spectra are properly normalized so they can be compared. The vertical dashed lines delimit the interval where the peak coming from the excess of 14.4 keV photon events, that would indicate a signal for $^{57}$Fe solar axions, is expected. As stated earlier, the TPC energy resolution ($\sigma$) at the 14.4 keV peak is 1.77 keV. Therefore, we regarded this $14.4 \pm 3.5$\nolinebreak[4] keV interval as the $\pm 2\sigma$ (95.45\%) signal region. To extract the signal, we subtracted the background spectrum from the tracking one, and the resulting signal spectrum can be seen on the right side in figure~\ref{fig:spectra}. The number of detected photon events in the signal region is $-71\pm57$. Consequently, only an upper limit on any axion signal expected in this energy range could have been set.

 Since the energy resolution of the TPC detector does not allow us to look at the very narrow 14.4 keV line alone, one also picks up a fairly broad part of the Primakoff solar axion spectrum which could mask the yield of $^{57}$Fe solar axions. This means that our method will not be able to identify their contribution for certain combinations of axion-photon and axion-nucleon coupling constant values. To exclude cases where the $\mbox{11-18}$~keV tail of the Primakoff flux exceeds the $^{57}$Fe solar axion flux in the measured signal spectrum, we set a constraint on the Primakoff axion emission by requiring
\begin{equation} \label{eq:Det1}
\int_{11~{\rm keV}}^{18~{\rm keV}}\frac{d\Phi_{\rm{a}}^{\rm{P}}(E_{\rm{a}})}{dE_{\rm{a}}}\,dE_{\rm{a}}<\Phi_{\rm{a}}\,.
\end{equation}
Using equations (\ref{eq:total_flux}) and (\ref{eq:Prim_flux}) this bound translates to
\begin{equation} \label{eq:Det2}
    g_{{\rm a}\gamma}<1.04\times 10^{-3}\,g_{\rm aN}^{\rm eff}\,\,\,\,\,\,{\rm GeV}^{-1}\,,
\end{equation}
and thus restricts the region of $g_{{\rm a}\gamma}-g_{\rm aN}^{\rm eff}$
parameter space in which our method, complemented by the use of the TPC detector, is strongly sensitive to $^{57}$Fe solar axions, as can be seen on the right side in figure~\ref{fig:upperlimits} (region below the ``Det" line). In this view, our analysis involves searching for the Gaussian-like energy spectrum of axion-induced photons,
\begin{equation}   \label{eq:gauss}
  N_{\rm{Fe}}(E)=\frac{1}{\sqrt{2\pi}\sigma}\,N_{\rm S}\,{\rm exp}\left[-\frac{(E-E_{\rm a})^2}{2\,\sigma^2}\right]\,,
\end{equation}
that describes the shape of the expected 14.4~keV axion peak in the measured signal spectrum. The analysis was performed by standard $\chi^2$ minimization. The energy resolution $\sigma=1.77$~keV and the position of the peak $E_{\rm a}=14.4$~keV were fixed during the fitting.
With $t=203$ hours of Sun tracking and $A=2\times 14.5$~cm$^{2}$ cross-sectional area of both magnet bores, the minimum of $\chi^2$/d.o.f.=11.47/13 corresponds to the number of signal events $N_{{\mathrm{S}}}=-42\pm27$. This result is consistent with zero, thus giving no evidence for $^{57}$Fe solar axions in our search. Therefore, an upper limit on the number of signal events, $N_{\mathrm{S}}<32$ counts at 95\% C.L., was obtained following the Bayesian approach~\cite{PDG08} by calculating the value of $N_{\mathrm{S}}$ that encompasses 95\% of the physically allowed part ($N_{\mathrm{S}}\ge 0$) of the Bayesian probability distribution. The corresponding fits are plotted in figure~\ref{fig:spectra} (right side).
The effect of the Primakoff axion tail on the result of our analysis was studied in the parameter space regions where the $^{57}$Fe solar axion contribution is comparable to or significantly less than the Primakoff one and discussed in section \ref{sec:results_discussion}.

\section{Results and discussion}                          \label{sec:results_discussion}

The number of photons expected to be detected by the TPC, coming from the magnet bores as a result of the conversion of $^{57}$Fe solar axions in the magnetic field, is
\begin{eqnarray} \label{eq:Nexpected}
N_{\rm{S}}=\int \frac{d\Phi_{\rm{a}}(E_{\rm{a}})}{dE_{\rm{a}}}\; P_{\rm{a\gamma}}(E_{\rm{a}})\;A\;t\; \epsilon(E_{\rm{a}}) \;dE_{\rm{a}}\,,
\end{eqnarray}
where $\epsilon(E_{\rm{a}})$ is the TPC detection efficiency for the photons of energy $E_{\rm{a}}$. We can rewrite the axion-photon conversion probability $P_{\rm{a\gamma}}(E_{\rm{a}})$, given by equation (\ref{eq:conversion_prob}), in a more convenient form as
\begin{eqnarray} \label{P_a_gamma_rewritten}
P_{\rm{a\gamma}}(E_{\rm{a}})=1.736\times 10^{3}\! \left( \frac{g_{\rm{a\gamma}}}{{\mathrm{GeV}}^{-1}}\right)^{2}\!\!
\left(  \frac{B}{9\;{\mathrm{T}}} \right)^{2}\!\!
\left(  \frac{L}{9.26\;{\mathrm{m}}} \right)^{2}\!
\frac{4}{q^{2}\:L^{2}}\, \sin^{2}\left( \frac{q\:L}{2}\right).\;\;\;\;\;
\end{eqnarray}
One should recall that the conversion probability depends implicitly on the axion mass $m_{\rm{a}}$ and the axion energy $E_{\rm{a}}$ through the axion-photon momentum difference $q=m_{\rm{a}}^{2}/2E_{\rm{a}}$. However, since the $^{57}$Fe solar axions are nearly monoenergetic, which is the result of their energy spectrum being very narrow (FWHM $\sim 5$ eV), it is a reasonable approximation that the conversion probability is constant over their energy range and equal to $P_{\rm{a\gamma}}\equiv P_{\rm{a\gamma}}(14.4$ keV). Hence, the equation\nolinebreak[4] (\ref{eq:Nexpected}) can be written as
\begin{eqnarray} \label{eq:Nexpected2}
N_{\rm{S}}= \Phi_{\rm{a}}\;P_{\rm{a\gamma}}\;A\;t\; \epsilon_{14.4} \,,
\end{eqnarray}
where $\epsilon_{14.4}=0.107$ is the detection efficiency of the TPC at 14.4 keV. Combining equations (\ref{P_a_gamma_rewritten}) and (\ref{eq:total_flux}), equation (\ref{eq:Nexpected2}) in our case becomes
\begin{eqnarray} \label{eq:Nexpected3}
N_{\rm{S}}=1.728\times 10^{33}\;(g_{\rm{aN}}^{\rm{eff}})^{2}\; \left( \frac{g_{\rm{a\gamma}}}{{\mathrm{GeV}}^{-1}}\right)^{2}\; \frac{4}{q^{2}\:L^{2}}\; \sin^{2}\left( \frac{q\:L}{2}\right),
\end{eqnarray}
where for the magnetic field we used its effective value of $8.83$~T, due to the fact that during 82\% of the data taking time the magnetic field was $8.79$~T, while for the rest of the time its value was $9$~T.

According to equation (\ref{eq:Nexpected3}), the upper limit on the number of signal events $N_{\rm{S}}<\nolinebreak[4]32$, resulting from the non-observation of the signal above back\-ground in our search for $^{57}$Fe solar axions, can be translated into the limit on the product of the parameters $g_{\rm{a\gamma}}$ and $g_{\rm{aN}}^{\rm{eff}}$ as\footnote{In the following we always mean $\left|g_{\rm{aN}}^{\rm{eff}}\right|$ when we write $g_{\rm{aN}}^{\rm{eff}}$.}
\begin{eqnarray} \label{eq:product_limit}
\frac{g_{\rm{a\gamma}}\,g_{\rm{aN}}^{\rm{eff}}}{{\mathrm{GeV}}^{-1}}<1.36\times 10^{-16}
\;\sqrt{\frac{q^{2}\:L^{2}}{4\,\sin^{2}\left( q\,L/2\right) }}\;\;\;\; (95\%\;{\mathrm{C.L.}}).
\end{eqnarray}
\begin{figure*}[t]
  \begin{minipage}{0.49\textwidth}
    \centerline{\includegraphics[width=0.95\textwidth]{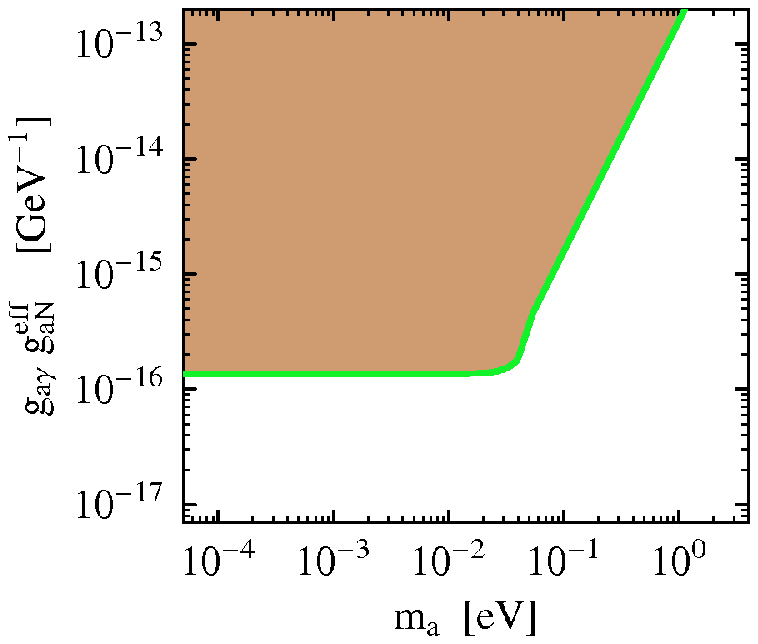}}
  \end{minipage}
  \begin{minipage}{0.49\textwidth}
    \centerline{\includegraphics[width=0.95\textwidth]{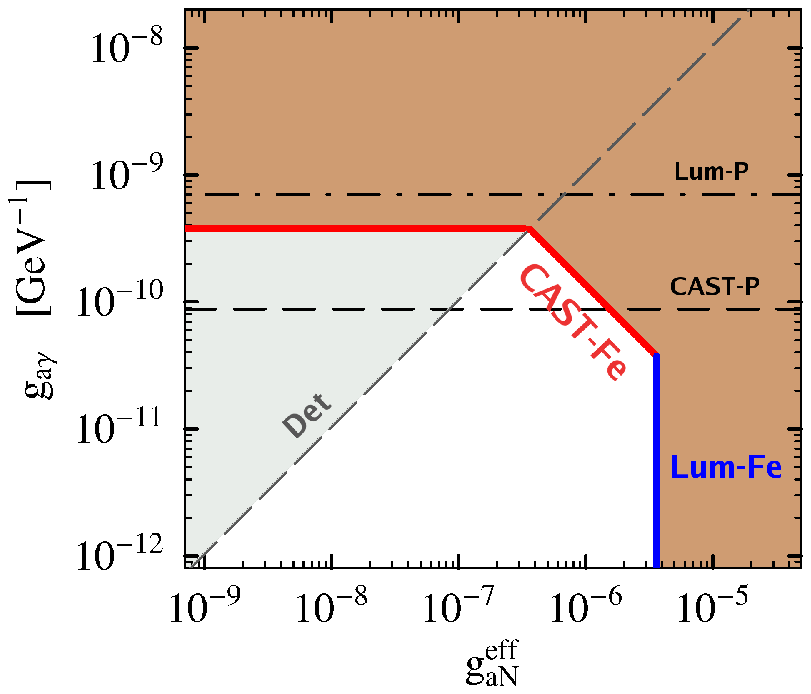}}
  \end{minipage}
 \caption{ Left: The upper limit (95\% C.L.) on the product $g_{\rm{a\gamma}}\,g_{\rm{aN}}^{\rm{eff}}$ as a function of the axion mass $m_{\rm{a}}$, imposed by the CAST's search for $^{57}$Fe solar axions. Right: The upper limit on $g_{\rm{a\gamma}}$ versus $g_{\rm{aN}}^{\rm{eff}}$, based on the relations $g_{\rm{a\gamma}}\,g_{\rm{aN}}^{\rm{eff}}< 1.36\times 10^{-16}$ GeV$^{-1}$ and $g_{\rm{a\gamma}}< 3.5\times 10^{-10}$ GeV$^{-1}$ for $m_{\rm{a}}\lesssim 0.03$ eV, is shown. The boundary denoted as Lum-Fe is a constraint due to the $^{57}$Fe solar axion luminosity, while the Det line is due to the detector resolution. This line divides the \mbox{$g_{\rm{a\gamma}}$--$g_{\rm{aN}}^{\rm{eff}}$ parameter} space into two regions where, roughly speaking, our method is dominantly sensitive to the $^{57}$Fe solar axions (below the line) and to the Primakoff axions (above), while in its proximity the sensitivities are comparable. Upper limits from the Primakoff solar axion luminosity (Lum-P)~\cite{Gondolo09} and CAST's search for the Primakoff solar axions (CAST-P)~\cite{Andri07}, that rely solely on $g_{\rm{a\gamma}}$, are also displayed for comparisons.
 \label{fig:upperlimits}}
\end{figure*}
We can consider this model-independent limit to be a function of the axion mass $m_{\rm{a}}$, as shown on the left side in figure~\ref{fig:upperlimits}. In the mass range $m_{\rm{a}}\lesssim 0.03$ eV, where the axion-photon conversion process is coherent and has maximum probability, the limit is mass-independent and its value (95\% C.L.) is
\begin{eqnarray} \label{eq:coherent_product_limit}
g_{\rm{a\gamma}}\,g_{\rm{aN}}^{\rm{eff}}<1.36\times 10^{-16}\;\;\; {\mathrm{{GeV}}^{-1}}\,.
\end{eqnarray}
For higher axion masses, an increase of the momentum mismatch $q$ causes a loss of the coherence of the axion-to-photon conversion, and suppresses the conversion probability due to the factor $\sin^{2}(x)/x^{2}$ in equation (\ref{eq:conversion_prob}), with  $x\equiv q L/2$. As a consequence, the sensitivity of the experiment to 14.4 keV axions diminishes rapidly with the increase of axion mass above $\sim$0.03 eV, thus providing weaker, mass-dependent limit, as can be seen from the plot in figure~\ref{fig:upperlimits} (left side).

The limit from equation (\ref{eq:coherent_product_limit}) allowed us to set the constraint on the axion-photon coupling constant as a function of the parameter $g_{\rm{aN}}^{\rm{eff}}$ for axion masses $m_{\rm{a}}\lesssim 0.03$ eV as
\begin{eqnarray} \label{eq:CAST-Fe}
g_{\rm{a\gamma}}<\frac{1.36\times 10^{-16}}{g_{\rm{aN}}^{\rm{eff}}}\;\;\; {\mathrm{{GeV}}^{-1}}\,.
\end{eqnarray}
This limit, denoted as ``CAST-Fe", is shown on the right side in figure~\ref{fig:upperlimits}. The vertical bound labeled ``Lum-Fe" at $g_{\rm{aN}}^{\rm{eff}}=3.6\times 10^{-6}$ is due to the requirement that the axion emission from $^{57}$Fe nuclei in the Sun should not exceed 10\% of the solar photon luminosity, as was explained in section \ref{sec:axion_emission}. The method we used to search for $^{57}$Fe solar axions can be applied in the region of $g_{\rm{a\gamma}}$-- $g_{\rm{aN}}^{\rm{eff}}$ parameter space given by equation (\ref{eq:Det2}) (i.e. below the line denoted as ``Det"), where the $^{57}$Fe solar axion flux exceeds the tail of the Primakoff solar axion flux in the energy range of the expected $^{57}$Fe axion signal. In this region, the dark brown area is excluded due to the $^{57}$Fe solar axion luminosity constraint and the ``CAST-Fe" line given by equation (\ref{eq:CAST-Fe}), while the white area is allowed according to our search. Due to the relatively low energy resolution of the TPC detector, the CAST search for $^{57}$Fe solar axions is not significantly sensitive in the parameter space region (grey) above the ``Det" line, where the Primakoff axion contribution dominates, while in the proximity of this line we expect that these two contributions are comparable. To examine how the Primakoff axion tail affects our result
(\ref{eq:coherent_product_limit}), we performed two additional $\chi^2$ analysis of the experimental signal spectrum using as fit functions: i) $N_{\rm Fe}+ N_{\rm P}$, and ii) $N_{\rm P}$. Here $N_{\rm Fe}$ is
given by equations~(\ref{eq:gauss}) and (\ref{eq:Nexpected3}), while $N_{\rm P}$  corresponds to the expected spectrum of the Primakoff axion tail (\ref{eq:Prim_flux}) multiplied by the axion-photon conversion probability (\ref{P_a_gamma_rewritten}) and the detection efficiency. The results of these fits are shown in table~\ref{tab:fits}. The suppression of our method's sensitivity to the $^{57}$Fe axions in the region above the ``Det" line due to the Primakoff axions contribution resulted in the upper limit of $g_{\rm{a\gamma}}<3.5\times 10^{-10}$~GeV$^{-1}$ at 95\% C.L., which is displayed as a red horizontal line in figure~\ref{fig:upperlimits} (right side). Also, according to table~\ref{tab:fits}, the effect of the Primakoff axion tail on our upper limit $g_{\rm{a\gamma}}\,g_{\rm{aN}}^{\rm{eff}}<1.36\times 10^{-16}$~GeV$^{-1}$ is $\sim$ 10\% in the proximity of the ``Det" line, while the $^{57}$Fe axions affect the limit $g_{\rm{a\gamma}}<3.5\times 10^{-10}$~GeV$^{-1}$ for about 9\%.

\begin{table*}[t]
  \centering
  \footnotesize
  \caption{ Results of $\chi^2$ analysis in the regions of $g_{\rm{a\gamma}}$--$g_{\rm{aN}}^{\rm{eff}}$ parameter space where our method is sensitive to $^{57}$Fe axions, $^{57}$Fe+Primakoff axions, and Primakoff axions (from top to bottom). }
  \begin{tabular}{cccccccc}
    \\ \hline\hline
    \multicolumn{1}{c}{Test} & \multicolumn{1}{c}{($g_{\rm{a\gamma}}\,g_{\rm{aN}}^{\rm{eff}})^{2}_{\rm best\, fit}
         \pm \sigma$} &
    \multicolumn{1}{c}{($g^4_{{\rm a}\gamma})_{\rm best\, fit}
         \pm \sigma$} &
    \multicolumn{1}{c}{$\chi^2_{\rm min}$/d.o.f.} &
    \multicolumn{1}{c}{$g_{\rm{a\gamma}}\,g_{\rm{aN}}^{\rm{eff}}$ (95\% C.L.)} &
    \multicolumn{1}{c}{$g_{{\rm a}\gamma}$ (95\% C.L.)} \\ hypothesis & $(10^{-32}$ GeV$^{-2})$ &
    $(10^{-38}$ GeV$^{-4})$ & & $(10^{-16}$ GeV$^{-1})$ & $(10^{-10}$ GeV$^{-1})$\\ \hline
    $N_{\rm{Fe}}$ & $-2.4\pm 1.6$   &  & $ 11.47/13$ & 1.36 & \\
    $N_{\rm{Fe}} + N_{\rm{P}}$ & $-1.7\pm 1.7$   & $-0.5\pm 1.3$   & $11.44/12$ & 1.5  & 3.8\\
    $N_{\rm{P}}$  & & $-1.1\pm 1.1$  & 12.89/13 &  & 3.5\\
    \hline\hline
    \label{tab:fits}
  \end{tabular}
\end{table*}

\begin{figure*}[t]
\begin{center}
 \centerline{\includegraphics[width=0.6\textwidth]{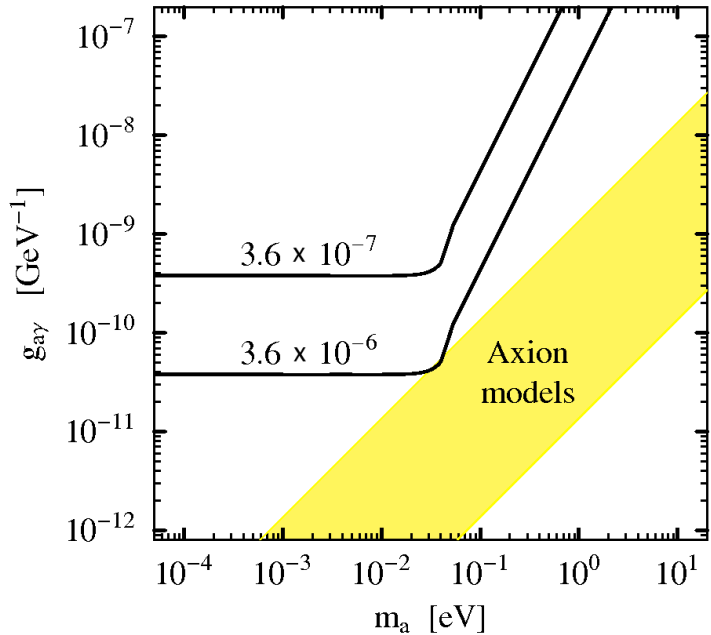}}
 \caption{
   Exclusion plot (95\% C.L.) in the $g_{\rm{a\gamma}}$ versus $m_{\rm{a}}$ plane. The limit achieved by our research is presented for two values of the parameter $g_{\rm{aN}}^{\rm{eff}}$ reflecting the luminosity and detection restrictions (from bottom to top, respectively). The yellow band represents typical theoretical axion models with
   $\left|E/N-1.95\right|$ in the range 0.07--7.
 \label{fig:fig5}}
 \end{center}
\end{figure*}

The experimental systematic uncertainties on the present limits were
studied. Regarding the background determination, the null hypothesis test
(in areas of the TPC detector where no signal is expected)
was used in order to estimate the systematic uncertainty induced by
possible uncontrolled dependencies of the background on time, position or
other experimental conditions. These effects were considered by
varying the background level until the null hypothesis test
yielded a result with a probability smaller than 5\%. If taken as an
uncertainty, this range corresponds to $\sim${\scriptsize $\begin{array}{c}
+8\% \\
-10\% \end{array}$} variation of the upper limits. As stated earlier, the deviation of our result due to the inclusion of possible contribution of the Primakoff axion tail in the fitting procedure was estimated to be less than +10\%. Other effects, such as uncertainties of the magnet parameters, are negligible, while the TPC efficiency affects the upper limit less than -2\%. Therefore, we estimated, using the quadratic sum of the individual contributions, that the overall effect of systematic uncertainties on our upper limit of $g_{\rm{a\gamma}}g_{\rm{aN}}^{\rm{eff}}$ is less than {\scriptsize $\begin{array}{c}
+13\% \\
-10\% \end{array}$}.

It is important to stress that the only axion properties we relied on in the entire procedure that led us to equation (\ref{eq:product_limit}) were that its couplings to photons and nucleons have a general form given by the Lagrangians in equations (\ref{eq:axion_nucleon_L}) and (\ref{eq:axion_photon_lagrangian}). We did not use any specific details regarding the coupling constants $g_{\rm{a\gamma}}$, $g_{\rm{aN}}^{0}$ and $g_{\rm{aN}}^{3}$ from any of the axion models.
Therefore, we can consider these coupling constants as free unknown parameters that characte\-rize the couplings of axions or general axion-like particles to two photons and a nucleon. This allowed us to use equation (\ref{eq:product_limit}) in order to set the upper limit on $g_{\rm{a\gamma}}$ as a function of $m_{\rm{a}}$ for various values of $g_{\rm{aN}}^{\rm{eff}}$. Figure~\ref{fig:fig5} shows the exclusion plots of $g_{\rm{a\gamma}}$ versus $m_{\rm{a}}$ obtained for two values of the parameter $g_{\rm{aN}}^{\rm{eff}}$ in comparison with the ``axion models band", i.e., the region of $g_{\rm{a\gamma}}$--\nolinebreak[4] $m_{\rm{a}}$ values expected from typical axion models with $|E/N-1.95|$ in the range 0.07--7.
The axion-nucleon couplings can vary from $3.6\times 10^{-6}$ to $3.6\times 10^{-7}$ reflecting constraints due to the $^{57}$Fe solar axion luminosity and detection sensitivity, respectively.
The presented contours do not enter the range of parameters, indicated by the yellow band in figure~\ref{fig:fig5}, that is predicted by plausible axion models to be the best-motivated region to search for axions. Thus, they should rather be considered as the limits on the two-photon coupling of axion-like particles that are somewhat lighter for a given interaction strength than it is expected for axions.

These two contours serve as an example to show how our result, given by the equation~(\ref{eq:product_limit}), can be used to scale the excluded region in the $g_{\rm{a\gamma}}$-- $m_{\rm{a}}$ parameter space for various choices of $g_{\rm{aN}}^{0}$ and $g_{\rm{aN}}^{3}$. In such a manner, it can be generally applied to impose the constraints on exotic light pseudoscalar particles that could be emitted in the nuclear magnetic transitions and couple to two photons. CAST has also performed a similar search for high-energy solar axions and ALPs from $^7$Li (0.478~MeV) and D($p$,$\gamma$)$^3$He (5.5~MeV) nuclear transitions, and the results are reported in \cite{And09}.

\section{Conclusion}                          \label{sec:conclusion}
The ongoing CAST experiment is primarily designed to search for
hadronic axions or more general ALPs of continuous energy spectrum, with
an average energy of 4.2~keV, that could be produced abundantly in the solar
core by the Primakoff conversion of thermal photons in the electric fields of
charged particles in the plasma. Since the reconversion of these particles
inside the CAST magnet bores would produce photons of the same energies, the
X-ray detectors used in CAST are optimized for the efficient detection of
photons in the 1-10~keV range. Here we explored the relation between the coupling constants of pseudoscalar particles that couple to a nucleon and to two photons by using the CAST setup during the Phase I to look for 14.4~keV monoenergetic solar axions and ALPs that may be emitted in the M1 nuclear transition of $^{57}$Fe. The signal we searched for, i.e., an excess of 14.4~keV X-rays when the magnet was pointing to the Sun was not found, and we set model-independent limits on the coupling constants of  $g_{\rm{a}\gamma}\:|-1.19\,g_{\rm{aN}}^{0}+
g_{\rm{aN}}^{3}|<1.36\times 10^{-16}$~GeV$^{-1}$ at
the 95\% confidence level. As a contrast to other experiments sensitive on the
$g_{{\rm a}\gamma}^2\, g_{\rm aN}^2$ couplings
\cite{Avignone88,Cha07,Borex08,Der09} that put some constraints in the  $\sim 10^2-10^6$~eV axion mass range, we explored the low mass region up to 0.03~eV.


\ack We thank CERN for hosting the experiment and for the contributions of
J.~P.~Bojon, F.~Cataneo, R.~Campagnolo, G.~Cipolla, F.~Chiusano,
M.~Delattre, A.~De~Rujula, F.~Formenti, M.~Genet, J.~N.~Joux, A.~Lippitsch, L.~Musa,
R.~De~Oliveira, A.~Onnela, J.~Pierlot, C.~Rosset, H.~Thiesen
and B.~Vullierme. We acknowledge support from NSERC (Canada), MSES (Croatia)
under the grant number 098-0982887-2872, CEA
(France), BMBF (Germany) under the grant numbers 05 CC2EEA/9 and 05
CC1RD1/0, the Virtuelles Institut f\"ur Dunkle Materie und Neutrinos --
VIDMAN (Germany), GSRT (Greece), RFFR (Russia), the Spanish Ministry of Science and Innovation (MICINN) under grants FPA2004-00973 and FPA2007-62833, NSF (USA),
US Department of Energy, NASA under the grant number NAG5-10842
and the helpful discussions within the network on direct dark matter
detection of the ILIAS integrating activity (Contract number:
RII3-CT-2003-506222).

\section*{References} 

\end{document}